\documentclass[twocolumn,prl,10pt,aps,floatfix]{revtex4-1}
\usepackage{graphicx}
\usepackage{amsmath}
\usepackage[colorlinks,linkcolor=blue,citecolor=blue,urlcolor=blue]{hyperref}
\usepackage{color}

\begin{document}
\title{Magnetothermal transport in the spin-1/2 easy-axis 
antiferromagnetic chain}  
\author{X. Zotos$^{1,2}$}
\affiliation{$^1$Department of Physics,
University of Crete, 70013 Heraklion, Greece}
\affiliation{$^2$
Max-Planck-Institut f\"ur Physik Komplexer Systeme, 01187 Dresden, Germany} 

\date{\today}

\begin{abstract}
By an exact analytical approach we study the magnetothermal transport in the 
spin-1/2 easy-axis Heisenberg model, in particular the thermal conductivity 
and spin Seebeck effect as a function of anisotropy, magnetic field and 
temperature. We stress a distinction between the common 
spin Seebeck effect with fixed boundary conditions and the one (intrinsic) 
with open boundary conditions. 
In the open boundary spin Seebeck effect 
we find exceptional features at the critical fields 
between the low field antiferromagnetic phase, the gapless one and 
the ferromagnetic at high fields. We further study the development 
of these features as a function of easy-axis anisotropy and temperature.
We point out the potential of these results to experimental studies in 
spin chain compounds, candidates for spin current generation in the field of 
spintronics.
\end{abstract}
\maketitle

\section{Introduction}
Over the last couple of decades the magnetic thermal transport has been 
established as a very efficient mode of thermal conduction \cite{hess}, 
next to the well known phononic and electronic ones.
Parallel to the search for evidence of 
ballistic thermal transport in quasi-one 
dimensional spin chain compounds described by the spin-1/2 Heisenberg 
Hamiltonian \cite{ballistic,znp}, 
numerous experimental studies focused on the effect of a
magnetic field on the thermal conductivity.
For instance, the spin-1/2 Copper Pyrazine Dinitrate \cite{pyrazine}, 
spin-one NENP \cite{nenp} and
ladder compounds \cite{ladder} were experimentally studied 
focusing on the interplay between the contributions of magnetic and phononic 
excitations and their mutual scattering.
Besides the magnetic thermal transport, only few recent studies 
were devoted to the spin Seebeck effect in quantum spin liquid systems, 
namely the generation of a spin current by a thermal gradient 
in a magnetic field. For instance,  
experimental studies on the Sr$_2$CoO$_3$ \cite{hirobe} 
compound with topological spinon excitations, 
CuGeO$_3$ with triplon excitations\cite{chen}, 
the spin-1/2 easy-axis antiferromagnet 
Pb$_2$V$_3$O$_9$ \cite{xing}  
and theoretical ones \cite{gros,sakai,meisner,psaroudaki}.

From a different perspective, 
the generation and control of spin currents is a central topic in the field 
of spintronics \cite{review}. 
In particular the spin Seebeck effect \cite{kikkawa}, 
has been extensively experimentally and 
theoretically studied in  a great variety of bulk magnetic compounds 
such as the 
ferrimagnetic YIG/Pt heterostructures and antiferromagnetic materials, 
e.g.  Cr$_2$O$_3$,~~~Fe$_2$O$_3$. 
Concerning the easy-axis antiferromagnetic materials,  
there is  experimental and theoretical 
interest and debate on the generated 
spin current sign change at the spin-flop transition.

Motivated by the abundance of Ising-like antiferromagnetic 
spin chain compounds and experimental studies over the years, 
e.g. on CsCoCl$_3$, CsCoBr$_3$, TMMC in the context of 
soliton -Villain- excitations\cite{mikeska}, 
phase diagram, spin dynamics,  quantum criticality of 
ACo$_2$V$_2$O$_8$ (A=Sr,Ba,Pb) \cite{co1,co2,co3}, 
we study the magnetothermal transport and in particular the 
spin Seebeck coefficient in the spin-1/2 easy-axis antiferromagnetic 
Heisenberg model. This study also serves as a bridge between spintronics 
studies in bulk materials and prototype magnetic systems.
We employ the Thermodynamic Bethe Ansatz (TBA) approach  
to analytically evaluate the relevant 
spin-energy current correlations within linear response theory.
We explore in particular, the sign of the spin current 
across the antiferromagnetic, gapless and ferromagnetic phases 
that characterize the Ising-like antiferromagnetic Heisenberg chain 
and the singular behavior at the critical fields. 

\section{Model and method}
We study the spin-1/2 antiferromagnetic Heisenberg model 
with easy-axis anisotropy, given by the Hamiltonian,
\begin{equation}
H=\sum_{l=1}^L J_{\perp} (
S^x_l S^x_{l+1} + S^y_l S^y_{l+1}) +
\Delta S^z_l S^z_{l+1} -h S^z_l,
\label{model}
\end{equation}
\noindent
where $S^{x,y,z}_l=\frac{1}{2}\sigma^{x,y,z}$, $\sigma^{x,y,z}$ Pauli 
spin matrices, $J_{\perp}>0$ is the easy-plane and $\Delta >0$ the easy-axis 
exchange interactions with $\Delta > J_{\perp}$ and $h$ the magnetic field.
Hereafter, we take $J_{\perp}=1$ as the unit of energy.

In linear response theory the spin and energy currents are related by 
the transport coeffients, $C_{ij}$,

\begin{equation}
\begin{pmatrix} \langle J^Q \rangle\\ \langle J^S \rangle \end{pmatrix} =
\begin{pmatrix} C_{QQ} & C_{QS} \\ C_{SQ} & C_{SS} \end{pmatrix}
\begin{pmatrix} -\nabla T \\ \nabla h \end{pmatrix}\,,
\label{MatrixEquation}
\end{equation}
where $C_{QQ}=\kappa_{QQ}$ ($C_{SS}=\sigma_{SS}$) is the heat (spin) 
conductivity and the thermal current $J^Q$ 
is related to the energy $J^E$ and spin 
current $J^S$ by, $J^Q=J^E-hJ^S$.
The coefficients $C_{ij}$ are given by the thermal average of 
time-dependent current-current correlation functions and it is 
straightforward to see that $C_{SQ}=\beta C_{QS}, (\beta=1/k_BT, k_B=1)$.
The real part of $C_{ij}(\omega)$ 
can be decomposed into a $\delta$-function 
at $\omega =0$ (the Drude weight) and a regular part:
\begin{equation}
Re(C_{ij}(\omega))=2\pi D_{ij}\delta(\omega)+
C^{reg}_{ij}(\omega).
\end{equation}
The spin-1/2 Heisenberg model is integrable 
by the Bethe ansatz method and transport is ballistic at finite 
magnetic fields \cite{znp}, with the energy current commuting with 
the Hamiltonian. Thus the magnetothermal coefficients are given 
by the Drude weights $D_{ij}$. 
We should note that, in view of experiments, 
this correspondence holds only if we assume 
the same relaxation rates for the magnetization and energy 
transport, $C_{ij}\sim D_{ij}\tau$ 
\cite{gros,sakai,meisner,psaroudaki}.
We will  consider the following situations:

\noindent
(i) $\langle J^S\rangle =0$, corresponding to a ''fixed boundary" system 
with spin accumulation, 
giving the thermal conductivity $\kappa$, 
\begin{equation}
\kappa=D_{QQ}-\beta\frac{D_{QS}^2}{D_{SS}}
=D_{EE}-\beta\frac{D_{ES}^2}{D_{SS}},
\end{equation}

\noindent 
${\rm MTC}=\beta\frac{D_{QS}^2}{D_{SS}}$ is the magnetothermal contribution 
and the spin Seebeck coefficient,
\begin{equation}
S=\frac{\nabla h}{\nabla T}=\beta(\frac{D_{ES}}{D_{SS}}-h);
\end{equation}

\noindent
(ii) $\nabla h=0$, corresponding to an ''open boundary" system \cite{md},
also referred to as intrinsic or bulk spin Seebeck effect \cite{intrinsic},
\begin{equation}
{\tilde S}=\frac{\langle J^S\rangle}{\langle J^Q \rangle}
=\beta\frac{D_{QS}}{D_{QQ}},
\label{ratio}
\end{equation}
\noindent
where $D_{QQ}=D_{EE}-2\beta h D_{ES} +\beta h^2D_{SS}$, 
$D_{QS}=D_{ES}-hD_{SS}$.

We evaluate the magnetothermal 
Drude weights in the framework of the  
TBA \cite{gaudin,takahashi,xz,xzth,doyon} approach.
In the easy-axis regime \cite{gaudin} the anisotropy is parametrized as 
$\Delta=\cosh\eta$ and in contrast to the easy-plane regime the 
Bethe ansatz solution is characterized by an infinite number 
of string excitations
with "particle" (and "hole") densities 
$\rho_j(x) (\rho_j^h(x)),~~j=1,\infty$, $x$ pseudomomenta.
$D_{SS}, D_{ES}, D_{EE}$, the specific heat $C$ and magnetic susceptibility,
are now given by the fairly standard TBA experssions,
\begin{eqnarray}
D_{SS}&=&\beta\sum_j\int_{-\pi}^{+\pi}dx r_j
n_j(1-n_j) (v^Q_j Q_j)^2
\nonumber\\
D_{ES}&=&\beta\sum_j\int_{-\pi}^{+\pi}dx r_j
n_j(1-n_j) (v^E_j E_j)(v^Q_j Q_j)
\nonumber\\
D_{EE}&=&\beta^2\sum_j\int_{-\pi}^{+\pi}dx r_j
n_j(1-n_j) (v^E_j E_j)^2
\nonumber\\
C&=&\beta^2\sum_j\int_{-\pi}^{+\pi}dx r_j
n_j(1-n_j) (E_j)^2.
\nonumber\\
\chi&=&\beta\sum_j\int_{-\pi}^{+\pi}dx r_j
n_j(1-n_j) (Q_j)^2.
\label{d}
\end{eqnarray}
\noindent
The total densities $r_j=\rho_j+\rho_j^h$ are obtained from, 
\begin{equation}
r_j=(\rho_j+\rho_j^h)=\frac{1}{\pi}a_j-\sum_k T_{jk}\ast \rho_k,
\label{r}
\end{equation}

\noindent
where ''$\ast$" denotes convolution symbol

\begin{eqnarray*}
a_j(x)&=&
\frac{ \sinh(j\eta)}{\cosh(j\eta)-\cos(x)},
\nonumber\\
T_{jk}&=&(1-\delta_{jk})a_{|j-k|}+2a_{|j-k|+2}+
\nonumber\\
&&...+2a_{j+k-2}+a_{j+k},
\nonumber\\
f\ast g(x)&=&\frac{1}{2\pi}\int_{-\pi}^{+\pi} f(x-y)g(y)dy
\end{eqnarray*}
\noindent
and the occupation numbers $n_j=1/(1+e^{\beta \epsilon_j})$, 
from the thermal energies $\epsilon_j$,
\begin{equation}
\epsilon_j=\epsilon_j^{(0)}+T\sum_k T_{jk}\ast
\ln(1+e^{-\beta\epsilon_k})
\label{the}
\end{equation}
\noindent
where $\epsilon^{(0)}_j=-\sinh\eta \cdot a_j(x)+hj$ are the bare
excitation energies. 
The effective velocities are given by \cite{bertini},
\begin{equation}
v^E_j=-v^Q_j=\frac{1}{2\pi r_j}\cdot
\frac{\partial \epsilon_j}{\partial x},
\label{v}
\end{equation}
\noindent
the "dressed" charges $Q_j$ and energies $E_j$,
\begin{eqnarray}
Q_j&=&Q_j^{(0)}-\sum_k T_{jk}\ast (n_k Q_k),~~Q_j^{(0)}=j
\nonumber\\
E_j&=&\epsilon^{(0)}_j-\sum_k T_{jk}\ast (n_k E_k),
\label{dressed}
\end{eqnarray}

\noindent
and the magnetization,
\begin{equation}
\langle S^z\rangle=\frac{1}{2}
-\frac{1}{2}\int_{-\pi}^{+\pi}dx r_jn_jQ_j.
\end{equation}

In the $T\rightarrow 0$ limit, there are three different 
phases \cite{johnson,bertini}: 
(i) for $h<h_{c}=\sqrt{\Delta^2-1}\cdot {\rm Dn}(\pi)$ 
it is  gapped antiferromagnetic, (ii)
for $h_{c}< h< h_f=1+\Delta$ it is a gapless spin liquid and (iii) 
for $h> h_f$ it is gapped ferromagnetic (
${\rm Dn}(x)=\frac{1}{2}\sum_{j=-\infty}^{+\infty}\frac{e^{ijx}}{\cosh(jx)}$
is the elliptic Jacobi function).

As $\epsilon_1<0,~\epsilon_j >0$ for $j>1$ in the low field antiferromagnetic
phase, we find that eqs.(\ref{the},\ref{dressed}) 
solved numerically by iteration with 
a finite cutoff in the number of strings
show poor or no convergence. 
The same applies when the thermal energies $\epsilon_j$ are numerically 
evaluated by the formulation of Ref.\cite{gaudin} and furthermore
the evaluated quantities do not accurately satisfy ''dressing" relations,
as for instance eqs.(\ref{r},\ref{dressed}) 
should imply $ \int dx r_j n_j Q_j^{(0)}= \int dx a_j n_j Q_j$.

To resolve the  convergence and "dressing" issues, we 
transform and solve by iteration eqs.(\ref{the}) by rewriting 
the term $\ln(1+e^{-\beta\epsilon_1})=-\beta\epsilon_1+
\ln(1+e^{\beta\epsilon_1})$\cite{fz}. 
Fourier transforming the equation 
for $j=1$ ($\hat{...}$ denotes Fourier transform), 
\begin{eqnarray*}
\hat{\epsilon}_1=
\frac{ \hat{\epsilon}^{(0)}_1 }{ 1+\hat{T}_{11} }
&+&T\cdot \frac{ \hat{T}_{11} }{ 1+\hat{T}_{11} }\cdot 
{\hat{\ln(1+e^{\beta\epsilon_1}}) }
\nonumber\\
&+&T\cdot \sum_{k>1} 
\frac{ \hat{T}_{1k} }{ 1+\hat{T}_{11} }\cdot 
{\hat {\ln(1+e^{-\beta\epsilon_k}}) },
\end{eqnarray*}

\noindent
back transforming,
\begin{eqnarray*}
\tilde{\epsilon}_1=\tilde{\epsilon}^{(0)}_1
&+&T\cdot \tilde{T}_{11}\ast
\ln(1+e^{\beta\epsilon_1})
\nonumber\\
&+&T\cdot \sum_{k>1} \tilde{T}_{1k} \ast \ln(1+e^{-\beta\epsilon_k})
\end{eqnarray*}
\noindent
and repeating the substitution for $j>1$,
\noindent
we obtain zero effective thermal energies $\tilde{\epsilon}^{(0)}_j$, 
densities $\tilde{r}^{(0)}_j$ and charges $\tilde{Q}^{(0)}_j$, 

\begin{eqnarray}
\tilde{\epsilon}^{(0)}_1&=&-\sinh(\eta)\cdot {\rm Dn}(x)+\frac{h}{2},
~~~\tilde{\epsilon}^{(0)}_{j>1}=(j-1)h
\nonumber\\
\tilde{r}^{(0)}_1&=&{\rm Dn}(x),~~~\tilde{r}^{(0)}_{j>1}=0
\nonumber\\
\tilde{Q}^{(0)}_1&=&\frac{1}{2},~~~\tilde{Q}^{(0)}_{j>1}=j-1.
\end{eqnarray}
\noindent
Note that the obtained effective energies are identical to those 
obtained in the low temperature antiferromagnetic regime 
\cite{johnson,bertini}.

\section{Results}
\begin{figure}[!ht]
\begin{center}
\includegraphics[width=1\linewidth, angle=0]{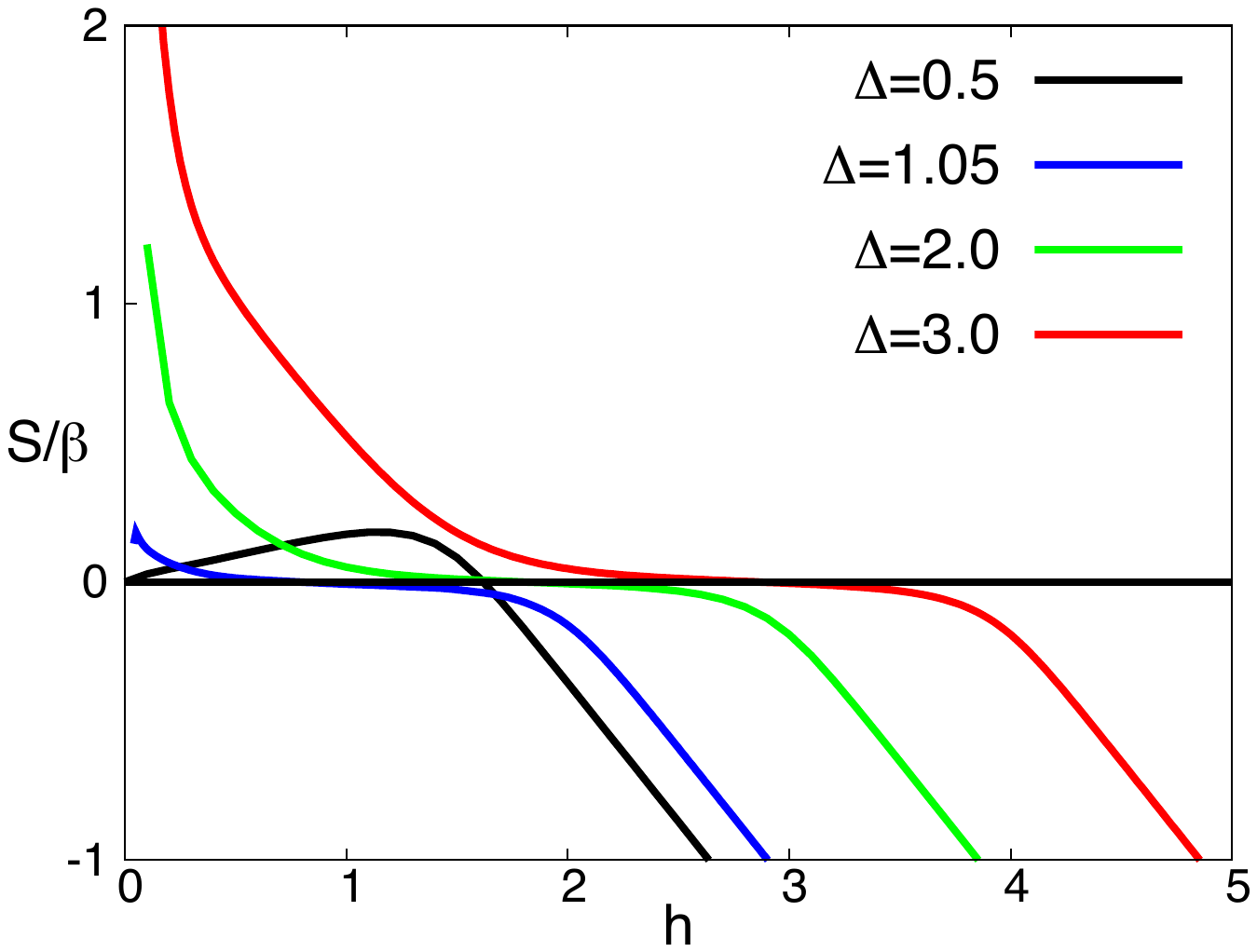}
\caption{Evolution of the spin Seebeck coefficient with anisotropy $\Delta$ 
as a function of magnetic field at temperature $T=0.1$.}
\label{fig1}
\end{center}
\end{figure}

In Fig. \ref{fig1} we show the spin Seebeck coefficient $S$ at low temperature
in the easy-axis and for comparison for $\Delta=0.5$ in the easy-plane 
regime. We find that $S$, in the gapless phase $h_c < h < h_f$,
decreases with decreasing anisotropy, 
diverges as $h\rightarrow 0$ and changes sign between the antiferromagnetic 
and ferromagnetic phases. In contrast to the easy-plane regime 
where the spin Drude weight $D_{SS}$ is finite and $S\rightarrow 0$ as 
$h\rightarrow 0$, the vanishing of $D_{SS},D_{ES}$ 
at $h=0$ for $\Delta > 1$ implies an ill-defined $S$. 
The results we find are consistent with the spin Seebeck coefficient 
evaluated at the isotropic limit \cite{sakai}. 
Of course, we expect physically the vanishing spin Drude weight $D_{SS}$ 
at $h=0$ to be replaced by a normal transport behavior, also at low 
temperatures, although this is still debated in studies 
focused in the high temperature limit \cite{1d}. 

\begin{figure}[!ht]
\begin{center}
\includegraphics[width=1\linewidth, angle=0]{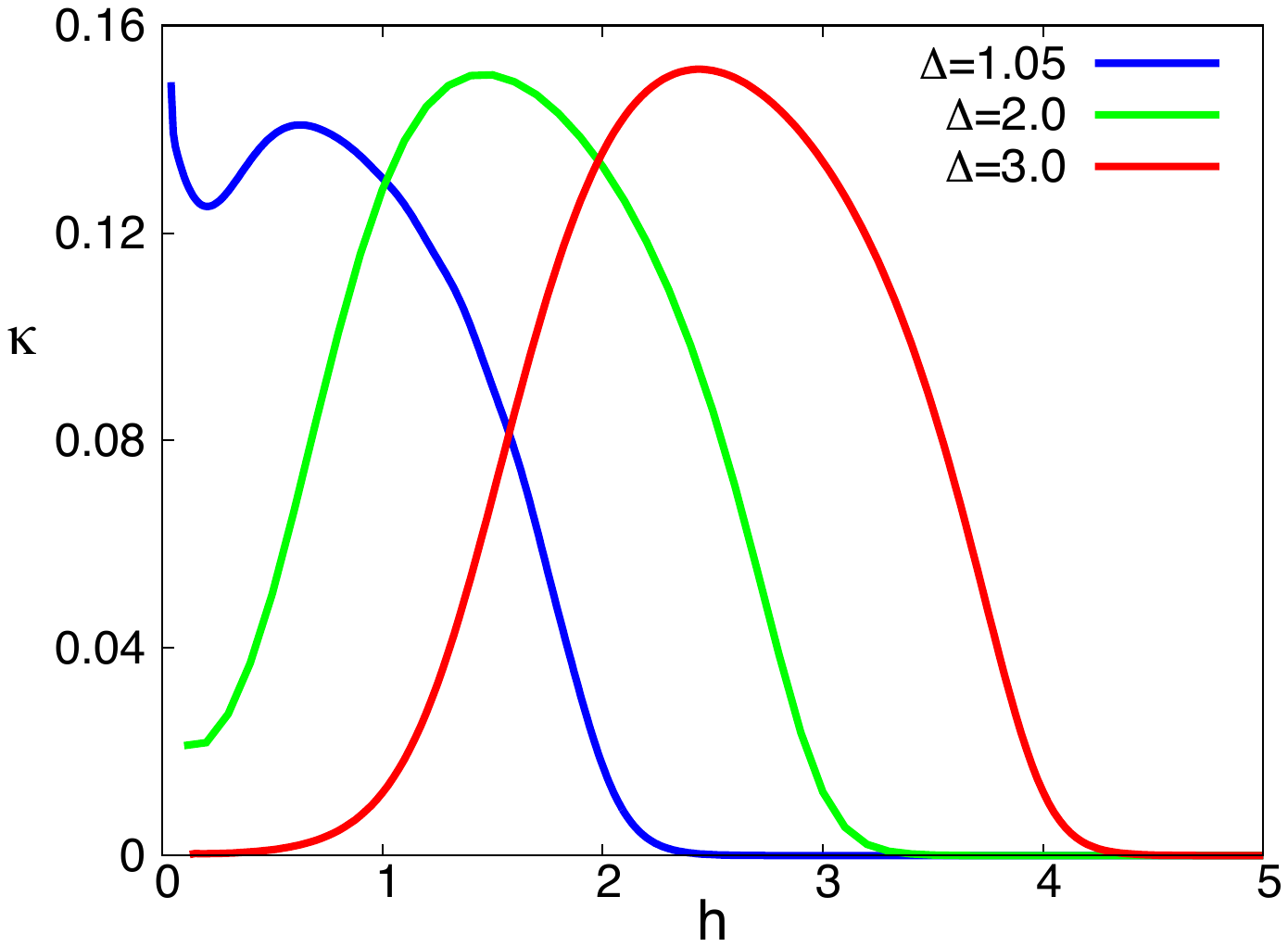}
\caption{Evolution of thermal conductivity with anisotropy $\Delta$ 
as a function of magnetic field at temperature $T=0.1$.}
\label{fig2}
\end{center}
\end{figure}

In Fig. \ref{fig2}, the thermal conductivity 
$\kappa$ is finite as $h\rightarrow 0$, although 
strongly suppressed in the gapped antiferromagnetic and ferromagnetic phase
for large anisotropy $\Delta$.
In particular $\kappa$ tends to a finite value as $h\rightarrow 0$ 
as the energy current commutes with the Hamiltonian and the thermal transport 
is purely ballistic over the whole phase diagram.

\begin{figure}[!ht]
\begin{center}
\includegraphics[width=1\linewidth, angle=0]{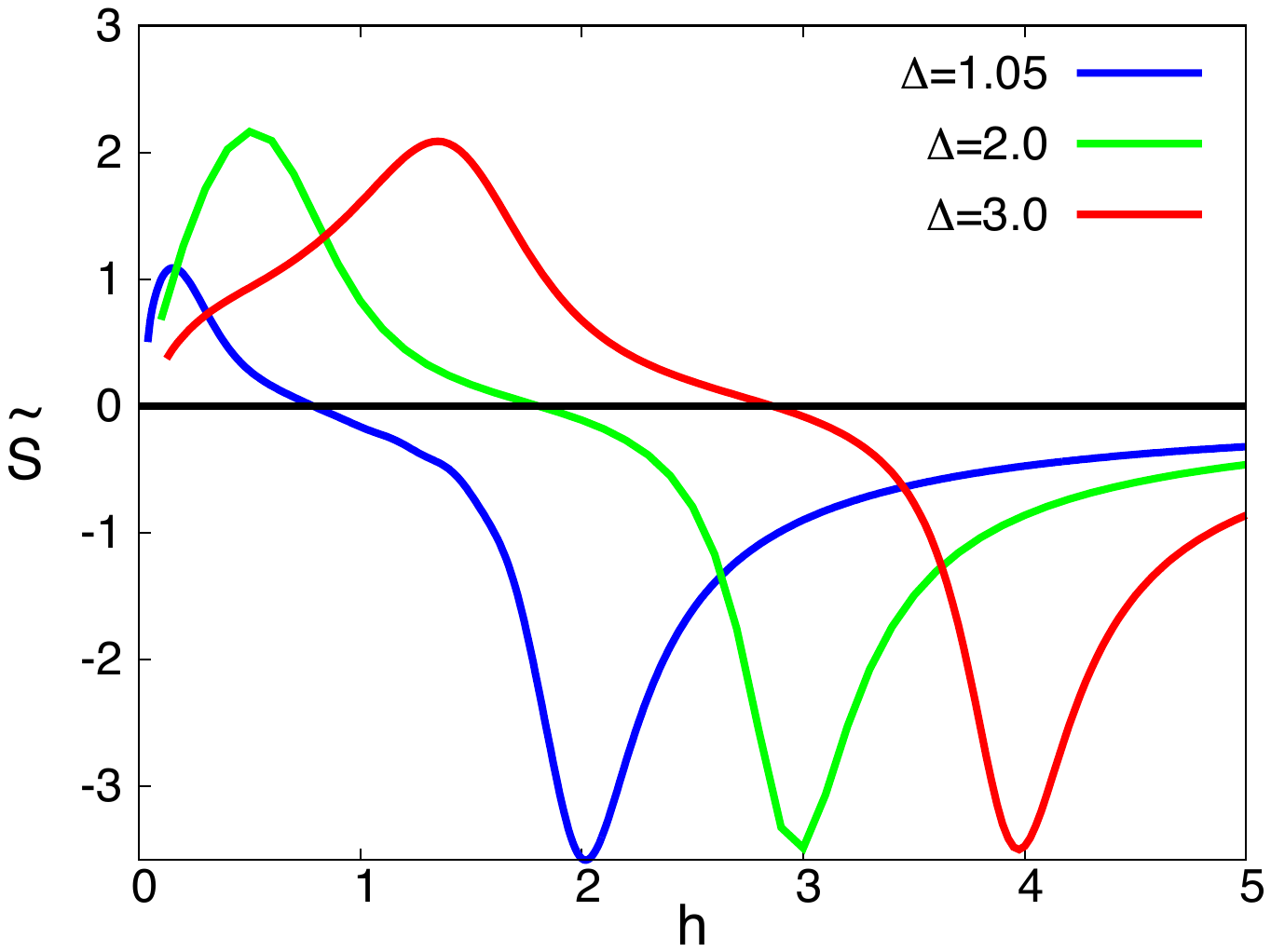}
\caption{Evolution of the ratio spin current to thermal current $\tilde S$ 
with anisotropy $\Delta$ 
as a function of magnetic field at temperature $T=0.1$.}
\label{fig3}
\end{center}
\end{figure}

In Fig. \ref{fig3} we show the ratio ${\tilde S}$ 
of the induced spin current to the thermal current. 
First, note that ${\tilde S}$ 
goes to zero as $h\rightarrow 0$ and there is a change 
of sign between the antiferromagnetic-gapless phase and 
the ferromagnetic one. However, pronounced features are developing 
at the critical fields $h_c,h_f$. It is remarkable that a similar behavior 
was found in a molecular dynamics and linear response study of the classical 
easy-axis Heisenberg model \cite{md}.  

\begin{figure}[!ht]
\begin{center}
\includegraphics[width=1\linewidth, angle=0]{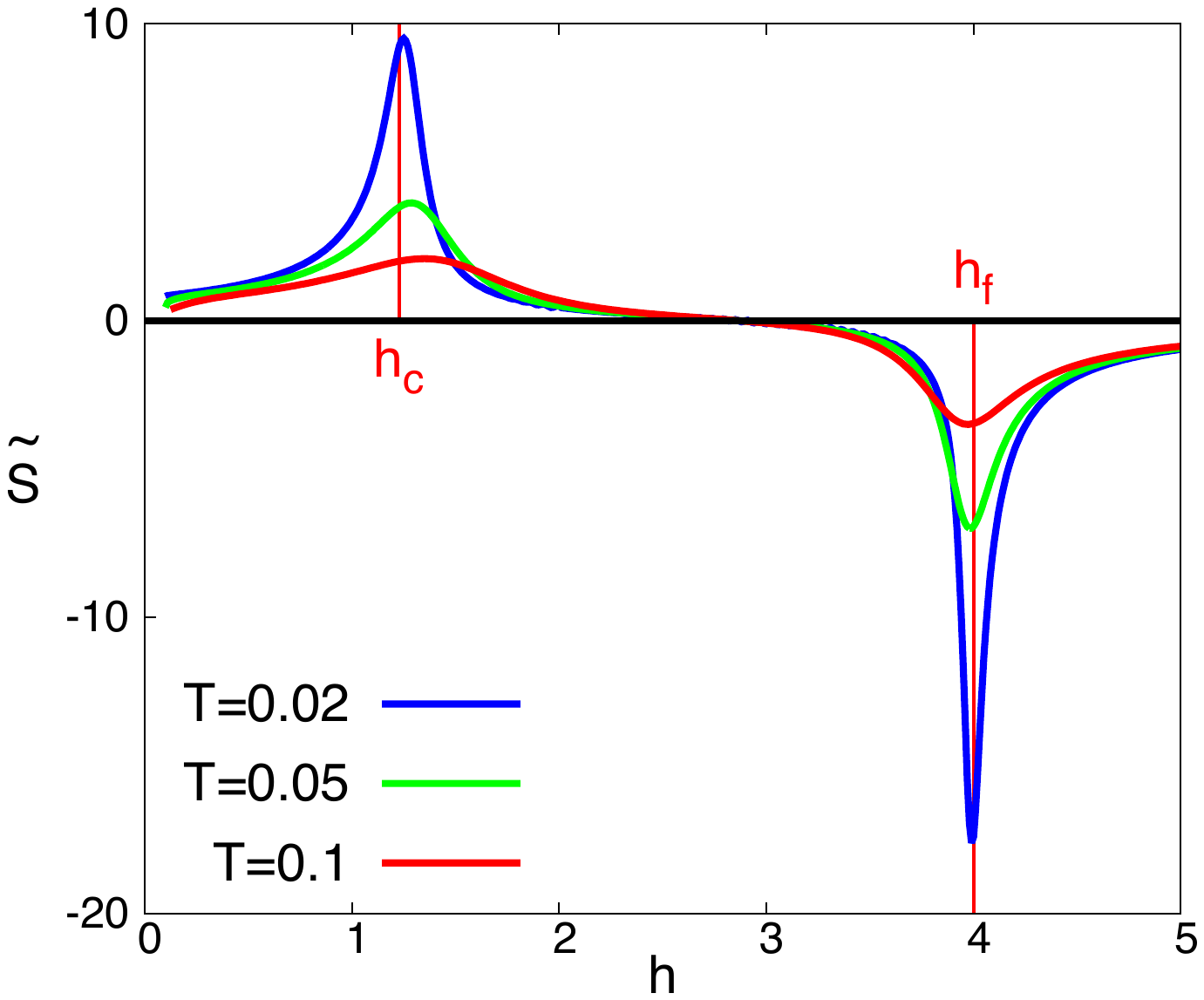}
\caption{Evolution of the ratio spin current to thermal current $\tilde S$ 
with temperature as a function of magnetic field at anisotropy $\Delta=3$.}
\label{fig4}
\end{center}
\end{figure}

\begin{figure}[!ht]
\begin{center}
\includegraphics[width=1\linewidth, angle=0]{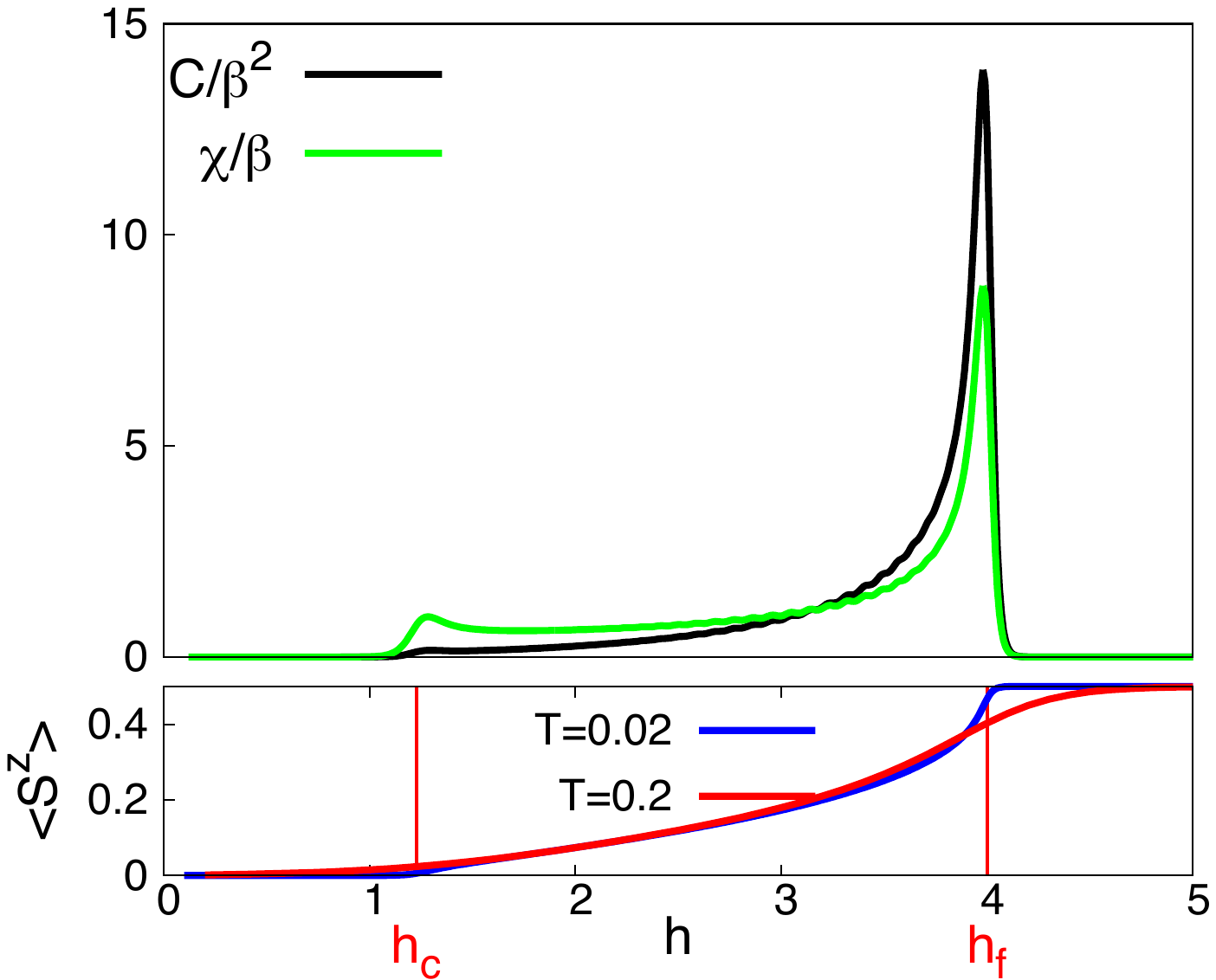}
\caption{Specific heat $C$ and magnetic susceptibility $\chi$ 
as a function of magnetic field at temperature $T=0.02$.}
\label{fig5}
\end{center}
\end{figure}

To further study the behavior at the phase transitions, 
in Fig. \ref{fig4} we show $\tilde S$ lowering the temperature 
at a rather large anisotropy $\Delta=3$ along with the 
critical fields $h_c, h_f$.
In the antiferromagnetic phase, the spin current vanishes for $h\rightarrow 0$ 
in contrast to $S$. At the transition between the gapless and 
the ferromagnetic phase there is a  
particularly pronounced peak at the critical field $h_f$.
This peak, as shown in Fig. \ref{fig5}, is related to the singular behavior 
of the specific heat and magnetic susceptibility at $h_f$ \cite{johnson2}.
It can be understood as the effect of the van Hove singularity 
in the density of states in the low density magnon system 
approaching the saturation field.
The peaks at the critical fields $h_c,h_f$ can be approximately 
described as Lorentzians of width proportional to the temperature.

In Fig. \ref{fig5}, we show the increase of 
magnetization as a function of magnetic field, from zero to saturation. 
Note that there is no exceptional singular 
behavior of the thermodynamic quantities at $h_c$, although there is one 
in ${\tilde S}$ as shown in Fig. \ref{fig4}.
We should note that the singularities in $\tilde S$ 
are related to the numerator $D_{QS}$ in Eq.\ref{ratio} which follows
a very similar pattern as a function of magnetic field (not shown).

\begin{figure}[!ht]
\begin{center}
\includegraphics[width=1\linewidth, angle=0]{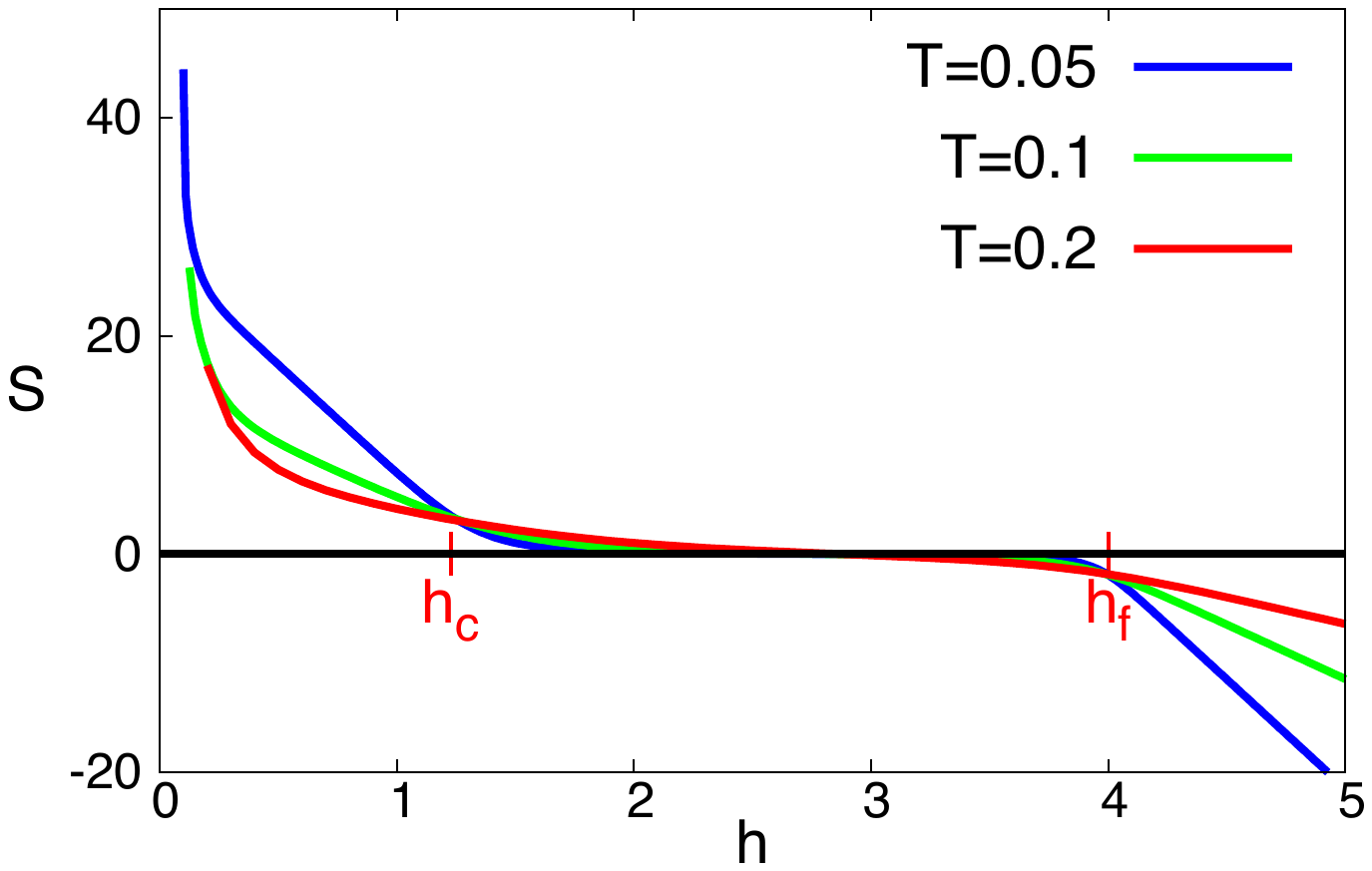}
\caption{Spin Seebeck coefficient as a function of magnetic field 
at different temperatures.}
\label{fig6}
\end{center}
\end{figure}

Fig. \ref{fig6} shows the spin Seebeck coefficient $S$ 
at different temperatures. It indicates a  
vanishing $S$ in the gapless phase as the temperature 
tends to zero, in accord 
with the induced spin current shown in Fig. \ref{fig4} and 
with calculation \cite{psaroudaki} 
in the gapless easy-plane ($\Delta < 1$) regime.
In contrast, in the antiferromagnetic and ferromagnetic gapped phases $S$ 
is finite and scales with $\beta$.

\begin{figure}[!th]
\begin{center}
\includegraphics[width=1\linewidth, angle=0]{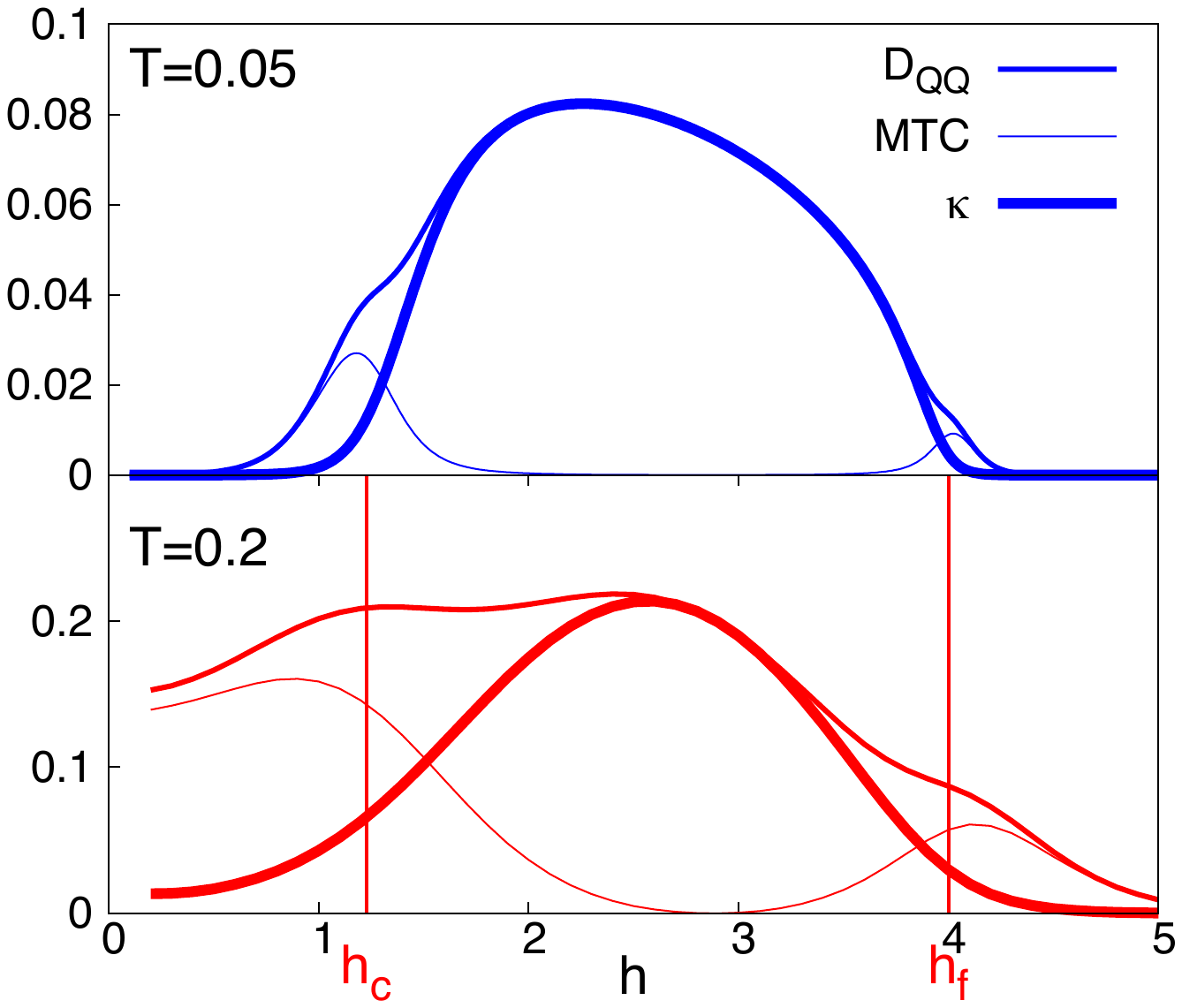}
\caption{Thermal conductivity and magnetothermal correction
as a function of magnetic field at $T=0.05$ and $T=0.2$ for $\Delta=3$.}
\label{fig7}
\end{center}
\end{figure}

Finally, in Fig. \ref{fig7}, we show the temperature dependence of the 
thermal conductivity $\kappa$ as a function of magnetic field and  
separately the contribution from the thermal current $D_{QQ}$ and 
the magnetothermal contribution MTC (an extesive discussion of the 
thermal Drude weight as 
a function of anisotropy and temperature was presented in \cite{sakaik}). 
Here, in contrast to the spin Seebeck 
coefficient, $\kappa$ is strongly  suppressed as expected 
in the antiferromagnetic and ferromagnetic gapped phases. 
We also find that the magnetothermal contribution is mostly 
relevant in the region of the critical fields at low temperatures.  

\section{Conclusions}
The main result of this work is that the 
open boundary (intrinsic, bulk)
spin Seebeck effect $\tilde S$ in the spin-1/2 easy-axis Heisenberg model 
shows exceptional features at the critical points of the phase diagram 
in contrast to the usual fixed boundary $S$ coefficient. 
In other words, the local spin current induced by a local thermal profile
has a very different magnetic field dependence from 
the accumulated magnetization in a fixed boundary system. 

Furthermore, the spin-1/2 model, although a quantum spin liquid,
shares the main features with the classical easy-axis Heisenberg model;
in particular, the singular behavior at the critical magnetic fields
and the sign change of the spin Seebeck coefficient 
between the antiferromagnetic and ferromagnetic phase.
But it also differs in the diverging 
spin Seebeck coefficient as 
$h\rightarrow 0$ that we might attribute to the integrability of the 
model. 

Quantum spin liquids are recently becoming candidates 
for spin current generation in the field of spintronics.
$\tilde S$ could be studied in experiments aimed at determining phase diagrams 
and detecting critical points, for instance in compounds as 
the spin-1/2 easy-axis Heisenberg chains  
ACo$_2$V$_2$O$_8$ (A=Sr,Ba,Pb) \cite{co1,co2,co3}.
The experimental challenge is to study  the ''open boundary" 
spin Seebeck effect, e.g. by local magnetothermal imaging \cite{landscape}.

Last but not least, further analytical study of this integrable model
should clarify the singular behavior of the induced  spin current 
in the vicinity of the critical fields at low temperatures.

\end{document}